Manipulation of Au nanoparticles using an electron probe: electron golf

Nan Jiang

Department of Physics, Arizona State University, Tempe AZ 85287-1504, USA

Abstract:

A tool for manipulation of specific nanoparticles is essential for nanofabrication. This paper describes a new class of nanoscale techniques for manipulating individual nanoparticles on an insulating substrate using a focused electron beam. The controlled displacement of 20 nm Au nanoparticles is demonstrated, along with the extension to other systems, by exploiting the fundamental principles of particle motion under intense electron irradiation, which we elucidate. Excitation and ionization by the focused electron beam within the substrate are shown to produce an electric field on the particle in this new mechanism. The strength and direction of the electric field, and thus the motion of the nanoparticle, can be precisely controlled by the current density, exposure time and position of the electron beam.

A novel nanotechnology, which can manipulate nanoparticles, is essential for the fabrication of quantum coherent structures and nanoscale devices. For the study of the underlying physics of the new affects arising from quantum confinement, in sity electron microscopy, in which many associated signals are available to probe the atomic and electronic structure (such as energy-loss spectroscopy, cathodoluminescence and atomic-resolution imaging and diffraction), is invaluable. The same electron beam which was used to construct the device may be used (at lower intensity) to probe the structure, at low temperature if needed.

Recently, the controlled motion of 1 – 2 nm Au particles or clusters on an amorphous carbon film has been achieved using a highly focused electron beam in the scanning transmission electron microscope (STEM) [1]. The driving forces are the interaction between the localized surface plasmons (the oscillating surface charges on the metallic nanoparticles) and the beam electron travelling in the proximity of the nanoparticle [2]. Due to the nature of the weak plasmonic response, only ultra-small Au nanoparticles (< 2 nm) can be moved, while larger ones (> 4 nm) are stationary [1]. The manipulation of larger nanoparticles (> 10 nm), dispersed in liquids, has also been reported recently. A 20 – 300 nm Al particle inside a molten Al-Si alloy could be moved using a focused electron beam in a transmission electron microscope (TEM) [3, 4]. Although it was considered to be similar to the optical trapping of dielectric spheres in liquids, the origin of the trapping forces for the Al particles was still unclear [5]. 10 nm Au nanoparticles dispersed in water in an environmental cell could also be trapped and steered using an electron beam in a TEM [6]. The trapping force was attributed to a negative pressure within the illuminated area, which was generated by rapid water evaporation under the electron beam [6]. The random rotation of a 20 nm CdSe nanoparticle floating on an amorphous C thin film has been observed in TEM [7]. It was suggested that the particle was trapped by a naturally formed, nanometer-sized 3-D Coulomb potential well, and rotated due to momentum transfer from the incident electrons.

In this letter, we introduce a new type of manipulation of nanoparticles: Au nanoparticles (~ 20 nm) move on an insulating solid substrate (silicate glass), driven by a stationary STEM probe. The probe size is about 0.2 – 0.3 nm in diameter, which is much smaller than the nanoparticle. The driving force for the displacement of an individual Au nanoparticle results from the induced electric field through excitation and ionization processes induced by the focused electron beam in the insulating substrate [8, 9]. By precisely controlling the probe current, probe position and exposure time, this method can move a particle on the surface to any desired location. Figure 1 shows an oblique view of the experimental geometry. The displacement of the Au particle was triggered by placing the highly focused electron beam (in "spot mode" in STEM), on one side of the particle, causing movement in the opposite direction, as illustrated in Fig. 1. This approach is different from all previously suggested methods for deliberately manipulating nanoparticles [1, 4, 6, 7]. First, the driving force is much stronger than the plasmonic forces [1]. Second, the moving particles are on a solid surface, rather than being dispersed in liquids [4, 6]. Third, and most importantly, the electron beam does not trap the moving particle as in the suggested electron tweezers' methods [1, 4, 6, 7], instead, it stays at the same position for the duration of the exposure (i.e. during the motion of the nanoparticle). The particle was evidently "kicked" from the initial to the final position by the electron probe, while the probe remained stationary (Fig. 1), as for a golf ball hit by a club.

The specimen used in this demonstration was a sodium borosilicate glass ($10Na_2O$-$20B_2O_3$-$70SiO_2$) containing 10 ~ 20 nm Au nanoparticles [10]. The TEM specimen was prepared by crushing the glass into a powder in acetone and mounting suspended pieces on a lacy carbon film covering a Cu grid. Therefore, the surface of the TEM specimen was pristine, and the exposed Au nanoparticles on the surface were uncontaminated. (A detailed description is given in the Supplementary Information.) The reason for choosing this glass was that the glass modifier $Na^+$ could be used as a "dye" to trace the induced electric field, and thus can be utilized to estimate the magnitude of electric forces acting on the moving particle. The controlled motion of Au nanoparticles was observed and analyzed in the Cornell VG HB501 100kV STEM, equipped with an ADF single-electron sensitivity detector and a parallel electron energy loss spectrometer. The probe size was about 2.2 Å, full width at half maximum, at which the saturated current of the electron probe was about 0.3 nA [11].

An example of controlled movement is shown in Fig. 2. Both BF and ADF images were recorded simultaneously by scanning the electron beam in an x-y raster, using both transmitted and scattered electrons, respectively. The contrast in the BF and ADF images are approximately complimentary to each other: the Au particles are dark in BF but bright in ADF. There are three Au nanoparticles in Fig. 2, and their relative positions are marked by a triangle. Fig. 2(a) was taken right after that the area was exposed to the electron beam and no particle had yet been illuminated by the stationary beam. Then the beam was stopped on particle A, with an exposure time of 5 seconds. The particle was intentionally displaced about 20 nm in the northwest direction, as shown in a subsequently acquired image (Fig. 2b); by placing the STEM probe on the lower right corner of the particle. The round-shaped featureless region centered at the stationary beam position, illustrated by the yellow circle in Fig. 2b, was also observed. This region, with a diameter of 40 – 50 nm, was created by the stationary electron beam, whose diameter was only about 0.2 – 0.3 nm. Such long-range interactions by a focused electron beam, which has often been observed in silicate glasses [12 – 15], is caused solely by the induced electric field [8, 9].

The mechanism of induced electric field motion in insulating materials by a STEM probe has been described previously [8], and is briefly explained as follows. In STEM, the electron beam, less than 0.5 nm in diameter, passes through a sample whose thickness is of the order of several tens or even hundreds of nanometers. Therefore, the beam-illuminated region can be considered as a nano-column or nano-rod [8]. This assumption is reasonable because the mean-free-paths (MFPs) of the emitted electrons (Auger and secondary electrons) are longer than the transverse dimension of the electron beam, and thus most of the emitted electrons travel out of the illumination column [8, 9]. These electrons are distributed randomly in the glass network over a much larger volume, due to their various MFPs. If the specimen is a good electron insulator, such as the silicate glass used in this work, the charge balance in the illuminated region cannot be immediately restored [16], resulting in a positively charged nano-rod [8, 9]. Consequently, the electric field produced by this charged nano-rod will eject the weakly bound cations, such as $Na^+$ [17, 18], and $Ge^{4+}$ in silicate [12, 15], toward nearby regions, forming a cation-less or even cation-free cylinder around the illuminated nano-rod. This cylinder is separated from the pristine matrix by a "wall" formed by accumulation of ejected cations. The lateral dimension of the cylinder is usually much larger than the probed column, and depends on beam current density and exposure time, due to the long-range interaction of the electric field (see supplementary materials and [8, 9]).

The projection of this electric field produced cylinder is clearly shown in the ADF image in Fig. 2b. According to our previous study, the contrast of this cylinder can be attributed to the ejection of Na, accompanied by a change in B coordination, from tetrahedral $[BO_4]$ to triangular $[BO_3]$ [17]. The lateral dimension of this cylinder is an indicator of the strength of the electric field. At the edge of the cylinder, the electric field drops to a strength that it is not strong enough to displace Na$^+$. Therefore, the dimension of the cylinder is determined by the migration of Na$^+$, driven by the induced electric field. To move an atom from one site to a nearest "equivalent site" in the [SiO$_4$] network, it requires a minimum of work to be done by the electric field which can at least overcome the activation energy $U_a$ for atom migration. So, the minimum strength of the induced electric field required to trigger atom migration is $E_{ind}^{Th} = U_a/aq$, in which, $a$ is the shortest distance between two minimum energy sites for the atom and $q$ is the electric charge of the atom. In this material, $U_a$(Na$^+$) ≈ 0.9 eV [19], and the Na – Na bond length is 0.33 nm [20]. So the shortest distance that a Na ion migrates should be on the nanometer scale, or less. Although the Na – O bond is ionic, it cannot be completely ionic, and thus the charge of Na should be a portion of the elementary charge +e (e = −1.602 × 10$^{-19}$ C). From these parameters, $a$ < 1 nm, $q$ < |e| and $U_a$ ~ 1.0 eV, we can estimate $E_{ind}^{Th}$ ~ 1.0 V/nm for Na$^+$ migration. Under the assumption that the STEM probed nano-rod is uniformly charged positively, the electric field has approximately a cylindrical symmetry around the probed rod and its strength is $E_{ind} \propto 1/r$, where $r$ is the horizontal distance from the probed column [8, 9]. Since the radius of the electric field induced cylinder, $R$, is in a range of 20 ~ 25 nm (Fig. 2), the maximum electric field in the probed region (~ 0.2 nm in diameter) should be in order of 100 V/nm. This induced electric field can also act on the Au particle if the particle carries transient net charges or is polarized into a dipole.

We can exclude the mechanism of recoil associated with Bragg diffraction (which, at Bragg condition **g**, transfers momentum $h$**g** to the entire mass of the gold nanocrystal in direction **g**) by the observation that the direction of motion of the particle is always along a line between the beam position and the particle center and is unrelated to the crystallographic orientation of the particle. In addition, we can also exclude the mechanism of melting Au nanoparticles heated by the electron beam. If the observed long-range motion was caused by melting, the nanoparticle would move in a particular direction. This is because the driving forces could be either gravitational or wettability gradients [21]. However, in our experiments, the motion of Au nanoparticles has no preferred direction. The experimental evidence for this is given in the SI.

In conclusion, we find that Au nanoparticles on a glass substrate can be positioned using a finely focused electron beam with nanometer precision. The driving force for the particle motion is an induced electric field. The method for moving a particle is to focus the electron beam on the side of the particle. The direction and distance of motion could be precisely controlled by the beam current density and exposure time. The significant of this study is that it suggests that nanoparticles may not be in an equilibrium state under an intense electron beam, and therefore, in some circumstances, the phenomena observed in TEM/STEM may not be representative of the pristine state of nano structures in normal matter. Future work, focused on more precise control will enable further improvements in the manipulation and atomic-scale modification of nanoparticles, thus opening up the field of nanomechanics and kinetics, which may be combined with in situ spectroscopy and atomic-resolution imaging.

Acknowledgement: This work is supported by DOE award DE-FG03-02ER45996. Acknowledgments are due to Prof. Jianrong Qiu of Zhejiang University for supplying the glass sample, Mr. Mick Thomas and Dr. Earl Kirkland of Cornell University for their technical support, Prof. John Silcox of Cornell University for his supervising of this research, and Prof. John Spence of Arizona State University for a critical reading of the manuscript.

Caption:

Figure 1 Schematic drawing showing displacement of a Au nanoparticle by focusing electron probe on a side of the particle. The displacement is toward other side of the particle. "Initial" and "final" indicate the positions before and after the Au nanoparticle moves, and **F** indicates the driving force acting on the particle parallel to the surface of substrate.

Figure 2 (a) BF (top row) and ADF (bottom) images showing three Au nanoparticles before the hit. (b) BF and ADF images showing the first move of an Au nanoparticle hit by the focused electron probe. The relative positions of three Au nanoparticles are marked by a triangle. The inset is the enlarged image of circled region in ADF image of (b).

Figure 1

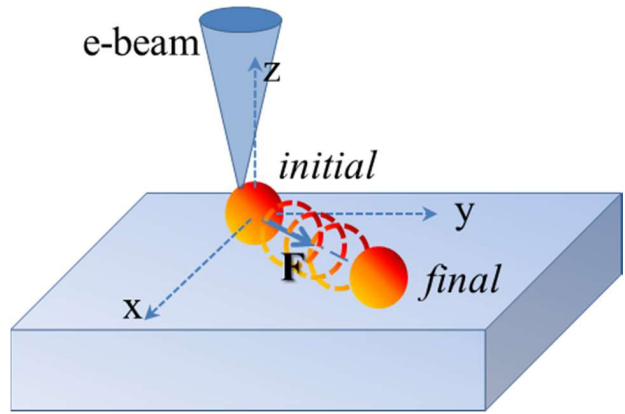

Figure 2

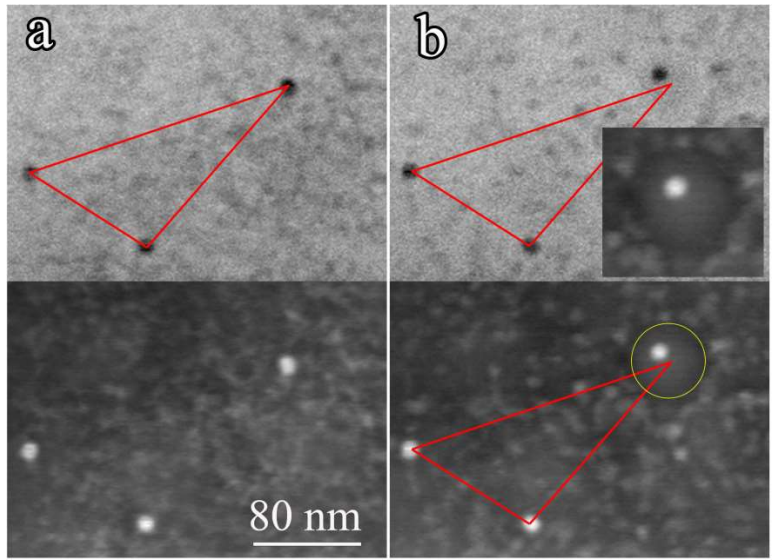